\documentclass[12pt,twoside]{article}
\usepackage{amsmath,amsfonts,amssymb}
\usepackage{latexsym}
\usepackage{dcolumn}
\usepackage{graphicx,epsfig}
\usepackage{amsthm}
\usepackage{hyperref}

\begin{document}

\begin{center}
{\Large \bf f(R) in Holographic and Agegraphic Dark Energy Models and the Generalized Uncertainty Principle}
\vglue 0.5cm
Barun Majumder\footnote{barunbasanta@iitgn.ac.in}
\vglue 0.6cm
{\small Indian Institute of Technology - Gandhinagar \\ Ahmedabad, Gujarat 382424\\ India}
\end{center}
\vspace{.1cm}

\begin{abstract} 
We studied a unified approach with the holographic, new agegraphic and the $f(R)$ dark energy model to construct the form of $f(R)$ which in general responsible for the curvature driven explanation of the very early inflation along with presently observed late time acceleration. We considered the generalized uncertainty principle in our approach which incorporated the corrections in the entropy area relation and thereby modified the energy densities for the cosmological dark energy models considered. We found that holographic and new agegraphic $f(R)$ gravity models can behave like phantom or quintessence models in the spatially flat FRW universe. We also found a distinct term in the form of $f(R)$ which goes as $R^{\frac{3}{2}}$ due to the consideration of the GUP modified energy densities. Although the presence of this term in the action can have its importance in explaining the early inflationary scenario but Capozziello {\it et.al.} recently showed that $f(R) \sim R^{\frac{3}{2}}$ leads to an accelerated expansion, {\it i.e.}, a negative value for the deceleration parameter $q$ which fit well with SNeIa and WMAP data.
\vspace{5mm}\newline Keywords: modified gravity, dark energy, generalized uncertainty principle
\end{abstract}
\vspace{1cm}

\section*{Introduction}

Observations of type IA supernovae confirms that our present universe is expanding at an accelerating rate \cite{f1}. Present observational cosmology has provided enough evidence in favour of the accelerated expansion of the universe \cite{f2,f3,f4}. Theoretical aid came in the form of exotic Dark Energy (DE) which can generate sufficient negative pressure and is believed to account nearly 70\% of present energy of the universe. Researchers in theoretical physics have proposed many DE models but they face problems while incorporating the history of the universe. The models generally have many free parameters and face serious constraints from observational data. Recent reviews \cite{f5add,f5,f6,f7,f7add} are useful for a brief knowledge of DE models.\par
The holographic DE is one of the promising DE model and the model is based on the holographic principle \cite{f8,f9,f10,f11,f12}. Bekenstein's entropy bound suggests that quantum field theory breaks down at large volumes. This can be reconciled by using a relation between UV and IR cutoffs such that $L^3 \Lambda^4 \leq L m_p^2$ where $m_p$ is the reduced Planck Mass ($m_p^{-2} = 8\pi G$). In this situation an effective local quantum field theory will give a good approximate description \cite{f13}. The holographic DE was first proposed in \cite{f14} following the line of \cite{f13} where the infrared cutoff is taken to be the size of event horizon for DE. The problem of cosmic coincidence can be resolved by the inflationary paradigm with minimal e-foldings in this model. Later this holographic DE was studied in detail by many authors \cite{f15,f16,f17,f18,f19,f20,f21,f22,f23,f24}. Clearly it can be mentioned that black hole entropy bound played an important role in the interpretation of holographic dark energy model. Various theories of quantum gravity (e.g., \cite{my25,my26,my27,my28,my29,my30,my31}) have predicted the following form for the entropy of a black hole:
\begin{equation}
S = \frac{A}{4l_p^2} + c_0 \ln \left( \frac{A}{4l_p^2} \right) + const.
\end{equation}
$c_0$ is a model dependent parameter and $l_p$ is the Planck length. Many researchers have expressed a vested interest in fixing $c_0$ (the coefficient of the subleading logarithmic term) \cite{my25}. Recent rigorous calculations of loop quantum gravity predicts the value of $c_0$ to be -1/2 \cite{my31}. A entropy corrected holographic DE model (ECHDE) was proposed recently in \cite{f48} where the inflation was driven by ECHDE. The curvature perturbation may be generated through the curvaton and the only requirement remain as $H \simeq const$ \cite{wei68a,wei68b}.\par
Another promising DE candidate is the agegraphic DE and was proposed in \cite{f49}. Considering the quantum fluctuations of spacetime K$\acute{a}$rolyh$\acute{a}$zy and his collaborators \cite{f50,f51,f52} argued that in Minkowski spacetime any distance $t$ cannot be known to a better accuracy than $\delta t \sim t_p^{\frac{2}{3}}t^{\frac{1}{3}}$, where $t_p$ is the reduced Planck time. Based on the arguments of K$\acute{a}$rolyh$\acute{a}$zy it can be shown that for Minkowski spacetime the energy density of metric fluctuations is given by $\rho_{\Lambda} \sim \frac{m_p^2}{t^2}$ \cite{f53,f54}. The agegraphic DE model considers spacetime and matter field fluctuations responsible for DE. If conformal time is considered in place of the age of the universe the model can describe the matter dominated epoch \cite{f55} with a natural solution to the coincidence problem \cite{f56} and is known as the new agegraphic DE model. The conformal time $eta$ is defined by $dt = a~d\eta$, where $t$ is the cosmic time and $a$ the scale factor. Many authors did some detailed study of this new agegraphic DE model \cite{f57,f58,f59,f60}. \par
Also we have other possible explanations for the cosmic acceleration, the different being the approach with $f(R)$ gravity, where $R$ is the scalar curvature. Other forms of $R$ along with $R$ in the Lagrangian can explain the observed acceleration without considering other additional components (the review \cite{f71} is useful). Among other existing theories $f(R)$ gravity models can be shown to be compatible with a matter dominated epoch transiting into an accelerating phase \cite{f88}. Also the forms of $f(R)$ with positive powers of curvature support the inflationary epoch and forms with negative powers of curvature serve as the effective DE responsible for cosmic acceleration and compatible with solar system experiments \cite{f89}. Also it is worth mentioning that these models face some challenges in the line of argument discussed in \cite{asph1,asph2,asphadd1,asphadd2,f103}. \par
The idea that the uncertainty principle could be affected by gravity was given by Mead \cite{new19}. In the regime when the gravity is strong enough, conventional Heisenberg uncertainty relation is no longer satisfactory (though approximately but perfectly valid in low gravity regimes). Later modified commutation relations between position and momenta commonly known as Generalized Uncertainty Principle (GUP) were given by candidate theories of quantum gravity (String Theory, Doubly Special Relativity Theory and Black Hole Physics) with the prediction of a minimum measurable length \cite{new20,new21,new22,new23,new24,new25,new26,new27,new28,new29}. Similar kind of commutation relation can also be found in the context of Polymer Quantization in terms of polymer mass scale \cite{new30}. Importance of the GUP can also be realized on the basis of simple gedanken experiments without any reference of a particular fundamental theory \cite{new27,new28}. So we can think the GUP as a model independent concept, ideally perfect for the study of black hole entropy. The authors in \cite{new31} proposed a GUP which is consistent with DSR theory, string theory and black hole physics. This is approximately covariant under DSR transformations but not Lorentz covariant \cite{new29}. With the GUP as proposed by the authors in \cite{new31} we can arrive at the corrected entropy-area relation for a black hole which can be written in the following expansive form \cite{new37,new}:
\begin{align}
\label{ent}
S  ~\simeq & ~\frac{A}{4l_p^2} + a~ \sqrt{\frac{A}{4l_p^2}} + b~ \ln \left(\frac{A}{4l_p^2}\right) \nonumber \\
& + \sum_{m=\frac{1}{2},\frac{3}{2},\ldots}^\infty d_m  \left(\frac{A}{4l_p^2}\right)^{-m} + 
\sum_{n=1,2,\ldots}^\infty c_n  \left(\frac{A}{4l_p^2}\right)^{-n} + \mathit{const.} ~~.
\end{align}
In this paper we will try to predict the form of $f(R)$ in holographic and new agegraphic DE models in the light of the generalized uncertainty principle \footnote{In \cite{ali} it has been argued that the holographic theory does not retain its good features by considering minimal length in Quantum Gravity. But here we will try to avoid the issue and hope to present the discussion in some future work.}. We will use eqn. (\ref{ent}) to calculate the energy density for the models. Later we will construct the form of $f(R)$ and the equation of state parameter $\omega$ for each of these DE models. Although an earlier attempt is present in the literature for the reconstruction of $f(R)$ \cite{f} but we will later conclude with a brief comparison of the results.

\section*{f(R) from Holographic DE model with GUP}

In $f(R)$ gravity the action is written as \cite{sud59,sud60,sud61}
\begin{equation}
S = \int \sqrt{-g}~ d^4x \left(\frac{R+f(R)}{16\pi G} + L_{matter}\right) ~~.
\end{equation}
The considerations lies in the fact that the higher order modifications of the Ricci curvature $R$ in the form of $R^2$ or $R_{\mu \nu}R^{\mu \nu}$ could give rise to inflation at the very early universe \footnote{The term $R_{\mu\nu}R^{\mu\nu}$ does not lead to any new kind of inflation different from that produced by the $R^2$ term since the combination $R_{\mu\nu}R^{\mu\nu}- \frac{1}{3}R^2$ does not contribute to the de Sitter solution at all.}. So this lead to a notion that whether inverse powers in $R$ dominant in late time universe can give an explanation to the recent predicted acceleration of the universe. But this type of models face problems of stability \cite{sud11}. The variation of the action with respect to the metric gives the field equations as
\begin{equation}
R_{\mu \nu} - \frac{1}{2}R g_{\mu \nu} = 8 \pi G \left(T_{\mu \nu}^{(R)} + T_{\mu \nu}^{(m)}\right) ~~,
\end{equation}
where 
\begin{equation}
8 \pi G~ T_{\mu \nu}^{(R)} = \frac{1}{2} g_{\mu \nu} f(R) - R_{\mu \nu} f'(R) + \left(\nabla_{\mu}\nabla_{\nu} - g_{\mu \nu} \Box\right) f'(R)~~.
\end{equation}
Here $f(R) = \frac{\partial f(R)}{\partial R}$, $R_{\mu \nu}$ is the Ricci tensor and $T_{\mu \nu}^{(m)}$ is the energy momentum tensor of matter and $R$ denotes the curvature contribution. \par
For a spatially flat FRW universe the modified Friedmann equation can be written as
\begin{equation}
H^2 = \frac{8\pi G}{3} (\rho_m + \rho_R)
\end{equation}
and
\begin{equation}
2\dot{H} + 3H^2 = - 8\pi G (p_m +p_R)
\end{equation}
where 
\begin{equation}
\label{maineq}
\rho_R = \frac{1}{8\pi G} \left[-\frac{1}{2}f(R) + 3(\dot{H} + H^2)f'(R) - 18(4H^2\dot{H} + H\ddot{H}) f''(R) \right] ~~,
\end{equation}
\begin{align}
p_R = \frac{1}{8\pi G} &\bigg[\frac{1}{2}f(R) - (\dot{H} + 3H^2)f'(R) + 6(8H^2\dot{H} + 6H\ddot{H} + 4\dot{H}^2 
+ \stackrel{...}{H}) f''(R) \nonumber \\
& + 36(\ddot{H} + 4H\dot{H})^2 f'''(R) \bigg] ~~
\end{align}
and
\begin{equation}
\label{eqR}
R = 6(\dot{H} + 2H^2)~~.
\end{equation}
Here the Hubble parameter is $H=\frac{\dot{a}}{a}$ and overdot denotes derivative with respect to cosmic time $t$. We can show that the curvature contribution will have its own equation of state and it can be written as \cite{f94}
\begin{align}
\label{eos}
\omega_R &= \frac{p_R}{\rho_R} \nonumber \\
&= 1 - \frac{4\big[\dot{H}f'(R) + 3(3H\ddot{H} - 4H^2\dot{H} + 4\dot{H}^2 + \stackrel{...}{H})f''(R) + 18(\ddot{H} + 4H\dot{H})^2 f'''(R)\big]}{\big[f(R) - 6(\dot{H} + H^2)f'(R) + 36(4H^2\dot{H} + H\ddot{H})f''(R)\big]} ~~.
\end{align}
In $f(R)$ gravity theories usually we encounter three types of scale factors for accelerating and inflationary cosmological solutions. We will follow the details of \cite{f95,f}. Here we will study phantom, quintessence and deSitter scale factors which are given by
\begin{equation}
\label{mGB2}
a=\left\{
\begin{array}{ll} a_0~(t_s - t)^{-h}, ~~~~~~~~~~~~ t\leq t_s,~~~~~ h>0~~~~~~~~~~ \mbox{(phantom)} \\\\
a_0 ~t^h,~~~~~~~~~~~~~~~~~~~~~~ h>0~~~~~~~~~~~~~~~~~~~~~~~~ \mbox{(quintessence)} \\\\
a_0 ~e^{Ht},~~~~~~~~~~~~~~~~ H=\mbox{constant~~~~~~~~~~~~~~~~~~(deSitter)}
\end{array} \right. 
\end{equation}
With the phantom scale factor and eqn.(\ref{eqR}) we get
\begin{equation}
H = \left[\frac{h}{6(2h+1)}R\right]^{\frac{1}{2}}
\end{equation}
and also 
\begin{equation}
\dot{H} = \frac{H^2}{2} ~~.
\end{equation}
Recent observations constrain the value of $h$ for the phantom scale factor to be $-\infty > h \geq 7.81$ \cite{obser}.
Similarly with the quintessence scale factor and eqn.(\ref{eqR}) we get
\begin{equation}
\label{quin1}
H = \left[\frac{h}{6(2h-1)}R\right]^{\frac{1}{2}}
\end{equation}
and also 
\begin{equation}
\label{quin2}
\dot{H} = -\frac{H^2}{2} ~~.
\end{equation}
For this quintessence scale factor the value of $h$ is very close to unity \cite{obser}.
For deSitter solution we have $H=$ constant. This scale factor is used describe the early inflationary scenario. For this case we get
\begin{equation}
\label{conh}
H = \left(\frac{R}{12}\right)^{\frac{1}{2}} ~~.
\end{equation}
Now we will try to evaluate the form of $f(R)$ for each of the scale factor mentioned above in the light of the generalized uncertainty principle (GUP). For our purpose we need to solve eqn.(\ref{maineq}) and we borrow the energy density from the holographic and agegraphic dark energy models respectively.\par
Considering the leading order terms of eqn.(\ref{ent}) and following the arguments of \cite{f48,wei6} we can write the GUP motivated energy density for the holographic DE model as
\begin{equation}
\label{hd}
\rho_{\Lambda} = \frac{3n^2 m_p^2}{L^2} + \frac{a m_p}{L^3} + \frac{b}{L^4} \ln(L^2 m_p^2) + \frac{c}{L^4} ~~.
\end{equation}
Here $n,a,b$ and $c$ are constants and $L$ is the future event horizon. If $a=b=c=0$ we get the usual holographic DE model. Though $n$ is a constant but its value can be constrained from the latest observational data \cite{f97}. The future event horizon is defined as
\begin{equation}
L = a \int_t^{\infty} \frac{d t}{a} ~~.
\end{equation}
For the phantom scale factor the future event horizon is 
\begin{equation}
L = a \int_t^{t_s} \frac{d t}{a} = \frac{1}{h+1} \sqrt{\frac{6h(2h+1)}{R}}
\end{equation}
Putting the value of $L$ in eqn.(\ref{hd}) we get the form of energy density as
\begin{equation}
\label{rhoh}
\rho_{\Lambda} = \frac{3n^2 m_p^2 (h+1)^2}{6h(2h+1)}R + \frac{a m_p (h+1)^3}{(6h)^{\frac{3}{2}}(2h+1)^{\frac{3}{2}}} R^{\frac{3}{2}} + \frac{(h+1)^4 R^2}{(6h)^2(2h+1)^2}\left[b\ln\left\{\frac{6h(2h+1) m_p^2}{(h+1)^2 R} \right\} + c \right] ~~.
\end{equation}
Now eqn.(\ref{maineq}) can be written in terms of $R$ as
\begin{align}
R^2 f''(R) &- \frac{(h+1)}{2}Rf'(R) + \frac{(2h+1)}{2}f(R) = -\frac{n^2 (h+1)^2}{2h}R  -\frac{a}{m_p} \frac{(h+1)^3}{(6h)^{\frac{3}{2}}(2h+1)^{\frac{1}{2}}} R^{\frac{3}{2}} \nonumber \\
&+ \frac{b}{m_p^2} \frac{(h+1)^4}{(6h)^2(2h+1)} R^2 \ln [R] - \frac{b}{m_p^2} \frac{(h+1)^4}{(6h)^2(2h+1)} \ln \left[ \frac{m_p^2 6h(2h+1)}{(h+1)^2}\right] R^2 \nonumber \\
&+ \frac{c}{m_p^2} \frac{(h+1)^4}{(6h)^2(2h+1)} R^2 ~~,
\end{align}
where $m_p^2 = \frac{1}{8\pi G}$. This equation is a nonhomogeneous Euler differential equation and the solution can be written as
\begin{equation}
\label{fR1}
f(R) = C_1 R^q + C_2 R^r + \delta R + \alpha R^{\frac{3}{2}} + \beta R^2 + \gamma R^2 \ln [R] ~~,
\end{equation}
where
\begin{equation}
\label{set1}
\left\{\begin{array}{ll} 
q = \frac{1}{4} \left[3 + h + \sqrt{h^2 - 10h + 1}~\right] \\\\
r = \frac{1}{4} \left[3 + h - \sqrt{h^2 - 10h + 1}~\right] \\\\
\delta = - \frac{n^2 (h+1)^2}{h^2} \\\\
\alpha = - \frac{1}{m_p} \frac{4a (h+1)^3}{(h+2)(6h)^{\frac{3}{2}} (2h+1)^{\frac{1}{2}}} \\\\
\beta = \frac{1}{m_p^2} \left[- \frac{b~(10+3h-h^2)(h+1)^4}{162~h^2(h+2)(2h+1)} - \frac{b~(h+1)^4}{54~h^2(2h+1)} \ln \left\{ \frac{m_p^2 6~h(2h+1)}{(h+1)^2}\right\} +  \frac{c~(h+1)^4}{54~h^2(2h+1)} \right] \\\\
\gamma = \frac{b}{m_p^2}\frac{(h+1)^4}{54~h^2(2h+1)}
\end{array} \right. 
\end{equation}
with $C_{1,2}$ as integration constants whose value can be predicted by the boundary conditions. The boundary conditions being 
\begin{equation}
f(R) \vert_{R=R_0} =- 2R_0  ~~~~~~~~~~     \text{and}  ~~~~~~~~~~  f'(R)\vert_{R=R_0} \sim 0 ~~,
\end{equation}
where $R=R_0$ is the present value of $R$ which is a small constant. The value of $R_0$ is of the order of $(10^{-33}eV)^2$. If we apply the boundary conditions we get the values of $C_{1,2}$ as
\begin{align}
C_1  =&  \frac{R_0^{1-q}}{2(q-r)} \Big[2(2r - \delta +  r \delta) - \alpha \sqrt{R_0}~(3 - 2r) - 2\beta R_0~ (2 -r) - \nonumber \\
& 2\gamma R_0 ~(1 + 2 \ln [R_0] - r \ln [R_0]~) \Big] ~~
\end{align} 
and
\begin{align}
C_2  =&  -\frac{R_0^{1-r}}{2(q-r)} \Big[2(2q - \delta +  q \delta) - \alpha \sqrt{R_0}~(3 - 2q) - 2\beta R_0~ (2 -q) - \nonumber \\
& 2\gamma R_0 ~(1 + 2 \ln [R_0] - q \ln [R_0]~) \Big] ~~.
\end{align}
In general the equation of state of eqn.(\ref{eos}) will be a function of $H$ and hence time in this case and so it can explain the transition from quintessence ($\omega_R > -1$) to phantom dominated regime ($\omega_R < -1$) as predicted by recent observations \cite{f100,f101,f102}. If we see eqn.(\ref{fR1}) we can infer that there is a contribution from $R^{\frac{3}{2}}$. This is interesting from the fact that we can have contributions from fractional powers of $R$. We will discuss this in the later part of this study.\par
For the quintessence scale factor the future event horizon is at 
\begin{equation}
\label{lqs}
L = a \int_t^{\infty} \frac{d t}{a} = \frac{1}{h-1} \sqrt{\frac{6h(2h-1)}{R}}
\end{equation}
with the condition $h>1$. Putting the value of $L$ from eq.(\ref{lqs}) in eqn.(\ref{hd}) we get the form of energy density as
\begin{equation}
\label{rhoq}
\rho_{\Lambda} = \frac{n^2 m_p^2 (h-1)^2}{2h(2h-1)}R + \frac{a m_p (h-1)^3}{(6h)^{\frac{3}{2}}(2h-1)^{\frac{3}{2}}} R^{\frac{3}{2}} + \frac{(h-1)^4 R^2}{(6h)^2(2h-1)^2}\left[b\ln\left\{\frac{6h(2h-1) m_p^2}{(h-1)^2 R} \right\} + c \right] ~~.
\end{equation}
So for the quintessence scale factor eqn.(\ref{quin1} and \ref{quin2}) we can rewrite eqn.(\ref{maineq}) with eqn.(\ref{rhoq}) as 
\begin{align}
R^2 f''(R) &+ \frac{(h-1)}{2}Rf'(R) - \frac{(2h-1)}{2}f(R) = \frac{n^2 (h-1)^2}{2h}R  +\frac{a}{m_p} \frac{(h-1)^3}{(6h)^{\frac{3}{2}}(2h-1)^{\frac{1}{2}}} R^{\frac{3}{2}} \nonumber \\
&- \frac{b}{m_p^2} \frac{(h-1)^4}{(6h)^2(2h-1)} R^2 \ln [R] + \frac{b}{m_p^2} \frac{(h-1)^4}{(6h)^2(2h-1)} \ln \left[ \frac{m_p^2 6h(2h-1)}{(h-1)^2}\right] R^2 \nonumber \\
&+ \frac{c}{m_p^2} \frac{(h-1)^4}{(6h)^2(2h-1)} R^2 ~~,
\end{align}
where $m_p^2 = \frac{1}{8\pi G}$. Similarly like the phantom case the solution can be written as
\begin{equation}
\label{fR2}
f(R) = C_1 R^q + C_2 R^r + \delta R + \alpha R^{\frac{3}{2}} + \beta R^2 + \gamma R^2 \ln [R] ~~.
\end{equation}
where
\begin{equation}
\label{set2}
\left\{\begin{array}{ll} 
q = \frac{1}{4} \left[3 - h + \sqrt{h^2 + 10h + 1}~\right] \\\\
r = \frac{1}{4} \left[3 - h - \sqrt{h^2 + 10h + 1}~\right] \\\\
\delta = - \frac{n^2 (h-1)^2}{h^2} \\\\
\alpha =  \frac{2a}{3m_p} \frac{(h-1)^3}{h(2-h)(2h-1)^{\frac{1}{2}}} \\\\
\beta = \frac{1}{m_p^2} \left[ \frac{b~(10-3h-h^2)(h-1)^4}{162~h^2(2-h)(2h-1)} + \frac{b~(h-1)^4}{54~h^2(2h-1)} \ln \left\{ \frac{m_p^2 6~h(2h-1)}{(h-1)^2}\right\} +  \frac{c~(h-1)^4}{54~h^2(2h-1)} \right] \\\\
\gamma = - \frac{b}{m_p^2}\frac{(h-1)^4}{54~h^2(2h-1)}
\end{array} \right. 
\end{equation}
The boundary conditions will give
\begin{align}
C_1  =&  \frac{R_0^{1-q}}{2(q-r)} \Big[2(2r - \delta +  r \delta) - \alpha \sqrt{R_0}~(3 - 2r) - 2\beta R_0~ (2 -r) - \nonumber \\
& 2\gamma R_0 ~(1 + 2 \ln [R_0] - r \ln [R_0]~) \Big] ~~
\end{align} 
and
\begin{align}
C_2  =&  -\frac{R_0^{1-r}}{2(q-r)} \Big[2(2q - \delta +  q \delta) - \alpha \sqrt{R_0}~(3 - 2q) - 2\beta R_0~ (2 -q) - \nonumber \\
& 2\gamma R_0 ~(1 + 2 \ln [R_0] - q \ln [R_0]~) \Big] ~~.
\end{align}
For the scale factor in deSitter space $H$ is constant. The future event horizon is located at
\begin{equation}
L = a L = a \int_t^{\infty} \frac{d t}{a} = \frac{1}{H} = \sqrt{\frac{12}{R}} ~~,
\end{equation}
where $H$ is given by eqn.(\ref{conh}). So we can write the GUP motivated energy density from eqn.(\ref{hd}) as
\begin{equation}
\rho_{\Lambda} = \frac{n^2 m_p^2}{4} ~R + \frac{a ~m_p}{12^{\frac{3}{2}}} ~R^{\frac{3}{2}} + \frac{b}{144~ m_p^2}~R^2 \ln \left(\frac{12~m_p^2}{R}\right) + \frac{c}{144~ m_p^2} ~R^2 ~~.
\end{equation}
So eqn.(\ref{maineq}) takes the form
\begin{equation}
R f'(R) - 2 f(R) = \frac{4 \rho_{\Lambda}}{m_p^2} ~~.
\end{equation}
The solution of this equation can be written in the form
\begin{equation}
f(R) = -n^2 ~R + C_1~R^2 - \frac{a}{3\sqrt{3}~m_p}~R^{\frac{3}{2}} - \frac{b}{72~m_p^2}~R^2~\left\{\ln \left(\frac{12~m_p^2}{R}\right)\right\}^2 + \frac{c}{36~m_p^2}~R^2~\ln (R) ~~,
\end{equation}
where $C_1$ is the arbitrary integration constant to be fixed by boundary conditions. The GUP motivated terms in $f(R)$ are important for inflationary scenario. We have instances for the $R^2$ term in literature to explain early time inflation \cite{sud59} as a curvature driven phenomenon. Here we have a new term $R^{\frac{3}{2}}$ in $f(R)$ which can be important for curvature driven inflation. We will discuss more about this term later in the discussion section.

\section*{f(R) from New Agegraphic DE model with GUP}

With the corrections due to the generalized uncertainty principle to the entropy area relation we can frame the energy density of the new agegraphic DE model \cite{f48,wei6} as
\begin{equation}
\label{rho2}
\rho_{\Lambda} = \frac{3n^2~m_p^2}{\eta^2} + \frac{a~m_p}{\eta^3} + \frac{b}{\eta^4}~\ln~(\eta^2~m_p^2) + \frac{c}{\eta^4} ~~,
\end{equation}
where $\eta$ is the conformal time. $a=b=c=0$ will give back the usual new agegraphic DE model. The parameter $n$ is constrained by present observations and its best fit value is around $2.716_{-0.109}^{+0.111}$ with $1\sigma$ uncertainty \cite{f56}. The numerical factor $3n^2$ was introduced for a parameterization of some uncertainties such as the effect of curved spacetime (as the K$\acute{a}$rolyh$\acute{a}$zy relation considered only the metric quantum fluctuations of Minkowski spacetime), the species of
quantum fields in the universe etc. \par
For the phantom scale factor the conformal time can be evaluated as
\begin{equation}
\eta = \int_t^{t_s}~\frac{dt}{a} = \frac{1}{a_0(h+1)} \left[\frac{6h(2h+1)}{R}\right]^{\frac{h+1}{2}} ~~,~~~~~~~~~~~~h>0 ~~.
\end{equation}
Substituting this in eqn.(\ref{rho2}) we get
\begin{align}
\label{rhap}
\rho_{\Lambda} &= \frac{3~n^2~m_p^2 ~a_0^2 (h+1)^2}{(6h)^{h+1}(2h+1)^{h+1}}~R^{h+1} + \frac{a ~m_p ~a_0^3 (h+1)^3}{(6h)^{\frac{3h+3}{2}}(2h+1)^{\frac{3h+3}{2}}}~R^{\frac{3h+3}{2}} + \nonumber \\
& \left[ \frac{b~a_0^4 (h+1)^4}{(6h)^{2h+2}(2h+1)^{2h+2}} ~\ln \left\{\frac{(6h)^{h+1}(2h+1)^{h+1}m_p^2}{a_0^2(h+1)^2}\right\}
+ \frac{c ~a_0^4 (h+1)^4}{(6h)^{2h+2}(2h+1)^{2h+2}} \right]R^{2h+2}  \nonumber \\
& - \frac{b~a_0^4 (h+1)^4}{(6h)^{2h+2}(2h+1)^{2h+2}} ~R^{2h+2} ~\ln ~(R^{h+1}) ~~.
\end{align}
Solving the inhomogeneous Euler differential equation (\ref{maineq}) with eqn.(\ref{rhap}) we get the form of $f(R)$ as
\begin{equation}
f(R) = C_1~R^q~ + C_2~R^r~ + \delta ~R^{h+1} + \alpha ~R^{\frac{3}{2}(h+1)} + \beta ~R^{2h+2} + \gamma ~R^{2h+2}~\ln ~R^{h+1} ~~, 
\end{equation}
where
\begin{equation}
\label{set3}
\left\{\begin{array}{ll} 
q = \frac{1}{4} \left[3 + h + \sqrt{h^2 - 10h + 1}~\right] \\\\

r = \frac{1}{4} \left[3 + h - \sqrt{h^2 - 10h + 1}~\right] \\\\

\delta = - \frac{3n^2 a_0^2 (h+1)^2}{h(6h)^{h+1}(2h+1)^h} \left[\frac{288 + 3360 ~h + 14816~ h^2+31360~h^3 +33408~h^4 +17280~h^5 +3456~h^6 }{288 + 3504~h +16496~h^2 +38768~h^3 +49088~h^4 +33984~h^5 + 12096~h^6 + 1728~h^7}\right] \\\\

\alpha = - \frac{a~a_0^3 (h+1)^3}{m_p~(6h)^{\frac{3h+3}{2}}(2h+1)^{\frac{3h+1}{2}}}\left[\frac{576 +4128~h + 10624~h^2 + 12032~h^3 + 6144~h^4 + 1152~h^5}{288 + 3504~h +16496~h^2 +38768~h^3 +49088~h^4 +33984~h^5 +12096~h^6 + 1728~h^7}\right] \\\\

\beta =  \bigg\{\left[\frac{b~ a_0^4 (h+1)^4}{m_p^2 (6h)^{2h+2}(2h+1)^{2h+1}} \ln\left\{\frac{(6h)^{h+1}(2h+1)^{h+1}m_p^2}{a_0^2 (h+1)^2}\right\} + \frac{c ~a_0^4 (h+1)^4}{m_p^2 (6h)^{2h+2}(2h+1)^{2h+1}} \right] \times \\
~~~~~\left[\frac{320+2528~h+6432~h^2+7168~h^3+3616~h^4+672~h^5}{288 + 3504~h +16496~h^2 +38768~h^3 +49088~h^4 +33984~h^5 +12096~h^6 + 1728~h^7}\right]\bigg\} + \\
~~~\left\{\left[\frac{b~ a_0^4 (h+1)^4}{m_p^2 (6h)^{2h+2}(2h+1)^{2h+1}} \right] \left[\frac{192+1696~h+4960~h^2+5920~h^3+3072~h^4+576~h^5}{288 + 3504~h +16496~h^2 +38768~h^3 +49088~h^4 +33984~h^5 +12096~h^6 + 1728~h^7}\right]\right\} \\\\

\gamma = \frac{b~ a_0^4 (h+1)^4}{m_p^2 (6h)^{2h+2}(2h+1)^{2h+1}} \left[\frac{192+1696~h+4960~h^2+5920~h^3+3072~h^4+576~h^5}{288 + 3504~h +16496~h^2 +38768~h^3 +49088~h^4 +33984~h^5 +12096~h^6 + 1728~h^7}\right]~~.

\end{array} \right. 
\end{equation}
The boundary conditions will give
\begin{align}
C_1 =& \frac{R_0^{1-q}}{2(q-r)} \bigg[4r - \alpha(3+3h-2r)~R_0^{\frac{3h+1}{2}} - 2(2\beta +2h\beta - r\beta +\gamma +h\gamma)~R_0^{2h+1} \nonumber \\
& -2\delta (1 + h -2r)~R_0^h - 2\gamma (2+2h-r)R_0^{2h+1}\ln ~[R_0^{h+1}]\bigg]
\end{align}
and
\begin{align}
C_2 =& \frac{R_0^{1-r}}{2(r-q)} \bigg[4q - \alpha(3+3h-2q)~R_0^{\frac{3h+1}{2}} - 2(2\beta +2h\beta - q\beta +\gamma +h\gamma)~R_0^{2h+1} \nonumber \\
& -2\delta (1 + h -2q)~R_0^h - 2\gamma (2+2h-q)R_0^{2h+1}\ln ~[R_0^{h+1}]\bigg]
\end{align}
In general the equation of state of eqn.(\ref{eos}) will be a function of $H$ and hence time in this case and so it can explain the transition from quintessence ($\omega_R > -1$) to phantom dominated regime ($\omega_R < -1$) as predicted by recent observations \cite{f100,f101,f102}.
For the quintessence scale factor (\ref{mGB2}) the conformal time can be evaluated with eqns.(\ref{quin1}) and (\ref{quin2}) as
\begin{equation}
\eta = \int_0^{t}~\frac{dt}{a} = \frac{1}{a_0(1-h)} \left[\frac{6h(2h-1)}{R}\right]^{\frac{1-h}{2}} ~~,~~~~~~~~~~~~\frac{1}{2}<h<1 ~~.
\end{equation}
For a real finite conformal time it is necessary to have $\frac{1}{2}<h<1$. Substituting this in eqn.(\ref{rho2}) we get
\begin{align}
\label{rhaq}
\frac{\rho_{\Lambda}}{m_p^2} &= \frac{3~n^2 ~a_0^2 (1-h)^2}{(6h)^{1-h}(2h-1)^{1-h}}~R^{1-h} + \frac{a ~a_0^3 (1-h)^3}{m_p~(6h)^{\frac{3-3h}{2}}(2h-1)^{\frac{3-3h}{2}}}~R^{\frac{3-3h}{2}}  \nonumber \\
& + \frac{ a_0^4 (1-h)^4}{m_p^2~(6h)^{2-2h}(2h-1)^{2-2h}}\left[b~ \ln \left\{\frac{(6h)^{1-h}(2h-1)^{1-h}m_p^2}{a_0^2(1-h)^2}\right\}
+ c \right]R^{2-2h}  \nonumber \\
& - \frac{b~a_0^4 (1-h)^4}{m_p^2~(6h)^{2-2h}(2h-1)^{2-2h}} ~R^{2-2h} ~\ln ~(R^{1-h}) ~~.
\end{align}
Solving the inhomogeneous Euler differential equation (\ref{maineq}) with eqns.(\ref{rhaq}), (\ref{quin1}) and (\ref{quin2}) we get the form of $f(R)$ as
\begin{equation}
f(R) = C_1~R^q~ + C_2~R^r~ + \delta ~R^{1-h} + \alpha ~R^{\frac{3}{2}(1-h)} + \beta ~R^{2-2h} + \gamma ~R^{2-2h}~\ln ~R^{1-h} ~~, 
\end{equation}
where
\begin{equation}
\label{set4}
\left\{\begin{array}{ll} 
q = \frac{1}{4} \left[3 - h + \sqrt{h^2 + 10h + 1}~\right] \\\\

r = \frac{1}{4} \left[3 - h - \sqrt{h^2 + 10h + 1}~\right] \\\\

\delta =  \frac{3n^2 a_0^2 (1-h)^2}{h(6h)^{1-h}(2h-1)^{-h}} \left[\frac{288 - 3360 ~h + 14816~ h^2 - 31360~h^3 +33408~h^4 -17280~h^5 +3456~h^6 }{-288 + 3504~h -16496~h^2 +38768~h^3 -49088~h^4 +33984~h^5 - 12096~h^6 + 1728~h^7}\right] \\\\

\alpha =  \frac{a~a_0^3 (1-h)^3}{m_p~(6h)^{\frac{3-3h}{2}}(2h-1)^{\frac{1-3h}{2}}}\left[\frac{-576 +4128~h - 10624~h^2 + 12032~h^3 - 6144~h^4 + 1152~h^5}{-288 + 3504~h -16496~h^2 +38768~h^3 -49088~h^4 +33984~h^5 -12096~h^6 + 1728~h^7}\right] \\\\

\beta =  \bigg\{\frac{ a_0^4 (1-h)^4}{m_p^2 (6h)^{2-2h}(2h-1)^{1-2h}}\left[b~ \ln\left\{\frac{(6h)^{1-h}(2h-1)^{1-h}m_p^2}{a_0^2 (1-h)^2}\right\} + c \right] \times \\
~~~~~\left[\frac{-192+1696~h-4960~h^2+5920~h^3-3072~h^4+576~h^5}{-288 + 3504~h -16496~h^2 +38768~h^3 -49088~h^4 +33984~h^5 -12096~h^6 + 1728~h^7}\right]\bigg\} - \\
~~~\left\{\left[\frac{b~ a_0^4 (1-h)^4}{m_p^2 (6h)^{2-2h}(2h-1)^{1-2h}} \right] \left[\frac{320-2528~h+6432~h^2-7168~h^3+3616~h^4-672~h^5}{-288 + 3504~h -16496~h^2 +38768~h^3 -49088~h^4 +33984~h^5 -12096~h^6 + 1728~h^7}\right]\right\} \\\\

\gamma =  \frac{-~b~ a_0^4 (1-h)^4}{m_p^2 (6h)^{2-2h}(2h-1)^{1-2h}} \left[\frac{-192 + 1696~h - 4960~h^2 + 5920~h^3 - 3072~h^4 + 576~h^5}{-288 + 3504~h -16496~h^2 +38768~h^3 -49088~h^4 +33984~h^5 -12096~h^6 + 1728~h^7}\right]~~.

\end{array} \right. 
\end{equation}
The boundary conditions $f(R) \vert_{R=R_0} =- 2R_0$ and $f'(R)\vert_{R=R_0} \sim 0$ will give
\begin{align}
C_1 =& \frac{R_0^{1-2h-q}}{2(q-r)} \big[4r~R_0^{2h} - \alpha (3 -3h-2r) ~R_0^{\frac{h+1}{2}} - 2(2\beta -2h\beta - r\beta +\gamma -h\gamma) ~R_0 \nonumber \\
& - 2\delta (1-h-r) ~R_0^h - 2\gamma (2 - 2h-r) ~R_0 ~\ln ~[R_0^{1-h}] \big]
\end{align}
and
\begin{align}
C_2 =& \frac{R_0^{1-2h-r}}{2(r-q)} \big[4q~R_0^{2h} - \alpha (3 -3h-2q) ~R_0^{\frac{h+1}{2}} - 2(2\beta -2h\beta - q\beta +\gamma -h\gamma) ~R_0 \nonumber \\
& - 2\delta (1-h-q) ~R_0^h - 2\gamma (2 - 2h-q) ~R_0 ~\ln ~[R_0^{1-h}] \big] ~~.
\end{align}
For the scale factor $a(t) = a_0~e^{Ht}$ with $H=constant$ (deSitter) we write the conformal time as 
\begin{equation}
\label{last1}
\eta = \int_0^{\infty} \frac{dt}{a} = \frac{1}{a_0H} = \sqrt{\frac{12}{a_0^2~R}} ~~.
\end{equation}
Here we have set the upper limit of the integration to $t\rightarrow \infty$ to express $\eta$ in terms of $R$. The relevant modification to the energy density (\ref{rho2}) will be
\begin{equation}
\label{last2}
\rho_{\Lambda} = \frac{3~n^2~m_p^2}{\eta^2} + \frac{a~m_p}{\eta^3} + \frac{b}{\eta^4} \ln~(\eta^2~m_p^2) + \frac{c}{\eta^4} ~~.
\end{equation}
The solution of eqn.(\ref{maineq}) with (\ref{last1}) and (\ref{last2}) yields the form of $f(R)$ as
\begin{equation}
f(R) = -n~a_0^2~R + C_1~R^2 - \frac{a~a_0^3}{3\sqrt{3}m_p}~R^{\frac{3}{2}} - \frac{b~a_0^4}{72~m_p^2}~R^2~\left[\ln \left(\frac{12~m_p^2}{a_0~R}\right)\right]^2  + \frac{c~a_0^4}{36~m_p^2}~R^2~\ln[R]~~.
\end{equation}
where $C_1$ is the integration constant to be fixed with boundary conditions. Here also like the holographic DE model we have a new term $R^{\frac{3}{2}}$ in $f(R)$ which can be important for curvature driven inflation.

\section*{Discussion}

In this study we considered the generalized uncertainty principle motivated forms of the holographic and the new agegraphic DE models to reconstruct the form of $f(R)$ suitable to explain the unification of early time inflation and late time acceleration. The idea that the Heisenberg uncertainty principle could be affected by gravity was given by Mead \cite{new19}. In the regime when the gravity is strong enough, conventional Heisenberg uncertainty relation is no longer satisfactory (though approximately but perfectly valid
in low gravity regimes). Modified commutation relations between position and momenta commonly known as the generalized
uncertainty principle (or GUP) were given by candidate theories of quantum gravity like string theory, doubly special relativity and black hole physics with the prediction of a minimum measurable length. Importance of the GUP can also be realized on the basis of simple gedanken experiments without any reference of a particular fundamental theory \cite{new27,new28}. So we can think the GUP as a model-independent concept suitable for the study of black hole entropy at least phenomenologically.\par 
According to the holographic principle the number of degrees of freedom of a bounded system should be finite and is related to the area of its boundary. As an application of the principle the upper bound of the entropy of the universe can be obtained. The total energy of a system of size $L$ should not exceed the mass of a black hole of the same size otherwise it would decay into a black hole. The saturation of the inequality means $\rho_{\Lambda} = \frac{3n^2~m_p^2}{L^2}$ where $m_p$ is the reduced Planck Mass ($m_p^{-2} = 8\pi G$). The UV cut-off is related to the vacuum energy and the IR cut-off is related to the large scale of the universe. The holographic dark energy scenario is viable if we set the IR cut-off by the future event horizon and also makes a concrete prediction about the equation of state of the DE \cite{f14}. On the other hand the new agegraphic DE model is based on the K$\acute{a}$rolyh$\acute{a}$zy relation which considers energy density of quantum fluctuations of the metric and matter in the universe. The energy density of the new agegraphic DE model has the same form as the holographic dark energy but the conformal time takes care of the IR cut-off instead of considering the future event horizon of the universe. The model not only account the observed value of DE in the universe but also predicts an accelerated expansion. Among various theoretical approaches to explain the present cosmic accelerated expansion of the universe only the holographic and the new agegraphic DE model is somehow based on the entropy-area relation. The entropy-area relation on the other hand can have quantum corrections through various approaches of quantum gravity.\par
As no single theoretical proposal for DE enjoys a pronounced supremacy over the others in terms of having a strong field theoretic support as well as being able to explain all the present observational data. This state of art explores another possibility of whether geometry in its own right could explain the presently observed accelerated expansion. The idea stems from the fact that higher order modifications of the Ricci curvature $R$ along with $R$ in the Einstein-Hilbert action could generate inflation in the very early universe. As the curvature is expected to fall off with the cosmic evolution it is then obvious whether inverse powers of $R$ in the action dominant during the later stages could drive a late time acceleration. In general this alternative theory is coined as $f(R)$ gravity.\par
In this paper we studied a unified approach with the holographic, new agegraphic and the $f(R)$ DE model to construct the form of $f(R)$ which in general responsible for the curvature driven explanation of the very early inflation along with presently observed late time acceleration. We considered the generalized uncertainty principle in our approach which incorporated the corrections in the entropy area relation which thereby modified the energy densities for the cosmological DE models considered here
\footnote{In the context of modified theories of gravity we should be cautious with the Wald entropy \cite{bamba8,bamba9} and not the Bekenstein-Hawking entropy. The Wald entropy is defined in terms of quantities on the Killing horizon and it depends on the variation of the Lagrangian density of the modified gravity theory with respect to the Riemann tensor. The Wald entropy is a local quantity and in $f(R)$ gravity it is given by $S_W = \frac{A~f'(R)}{4G}$ \cite{bamba10}. But here we have just reconstructed $f(R)$ from the energy densities of other DE models so we have not considered the Wald entropy.}.
We found that the GUP motivated holographic and new agegraphic $f(R)$ gravity models can behave like phantom or quintessence models in the spatially flat FRW universe. A similar study was also carried out by authors in \cite{f}. We reproduced all the result and conclusion of \cite{f} but in addition we also found a distinct term in the form of $f(R)$ which goes as $R^{\frac{3}{2}}$ due to the consideration of the GUP modified energy densities. This is really very interesting if we consider the phenomenological consequence of our study. Although the presence of this term in the action can have its importance for inflation but Capozziello {\it et.al.} \cite{capo1,capo2} introduced an action with $f(R) \sim R^m$ and showed that it leads to an accelerated expansion, {\it i.e.}, a negative value for the deceleration parameter $q$ for $m \approx \frac{3}{2}$ which fit well with SNeIa and WMAP data. Apart from the $R^{\frac{3}{2}}$ term we also found the other possible contributions of $R$ like $R^2$ and $R^2~\ln [R]$ which also have importance in the inflationary scenario. We should also mention here that in the latter case one needs not only quasi-exponential expansion but a metastable (i.e. slowly rolling) one. In $f(R)$ gravity this may occur only if $f(R)$ is close to $R^2$ over some range of $R$ \cite{refsug}.

\section*{Acknowledgements}
The author was partly supported by the Excellence-in-Research Fellowship of IIT Gandhinagar. The author would like to thank an anonymous referee for enlightening comments which immensely helped in improving the manuscript. 



\end{document}